\documentclass[10pt,a4paper,twoside]{article}
\usepackage{epsfig}
\usepackage{baltlat5}
\usepackage{wrapfig}
\begin{document}
\ \
\vspace{0.5mm}

\setcounter{page}{1}
\vspace{5mm}

\titlehead{Baltic Astronomy, vol.\ts 14, XXX--XXX, 2005.}

\titleb{SPECTROSCOPIC ANALYSIS OF SDB BINARIES \\ FROM THE
  SPY PROJECT}

\begin{authorl}
\authorb{C. Karl}{1}
\authorb{U. Heber}{1} 
% \authorb{S. Jeffery}{3} and
\authorb{R. Napiwotzki}{2} 
\authorb{S. Geier}{1} 
\end{authorl}

\begin{addressl}
\addressb{1}{Dr.\,Remeis-Sternwarte, Sternwartstra\ss e 7, 96049 Bamberg, Germany}
% \addressb{3}{Armagh Observatory, College Hill, Armagh BT61 9DG, Northern Ireland, UK}
\addressb{2}{Centre for Astrophysics Research, University of Hertfordshire, College Lane, Hatfield AL10 9AB, UK}
\end{addressl}

\submitb{Received 2005 April 1}

\begin{abstract}
In the course of our search for double degenerate binaries as potential
 progenitors of type Ia supernovae with the ESO VLT 
% (ESO\,-\,SN\,Ia progenitor survey\,-\,SPY)
 several new subdwarf B (sdB) binaries 
% with sdB primary components 
 were discovered.
In this paper, we present detailed analyses of six 
 radial velocity variable sdB stars.
Radial velocity curves have been measured.
From the mass functions we derive lower limits to the 
 masses of the unseen companions and we discuss their nature.
% We deduce the nature of the unseen companions and discuss the results.
In addition, stellar parameters like effective temperatures, 
 surface gravities and helium abundances were
 determined as well as metal abundances.

\end{abstract}

\vskip4mm


\resthead{Spectroscopic analysis of sdB binaries from SPY}{C.\,Karl~et.~al.}

\sectionb{\bf 1}{ \bf INTRODUCTION}\\
There is general consensus that the precursors
 of type Ia supernovae (SN Ia) are white dwarfs in close binary systems.
These white dwarfs accrete matter from their companions until a
 critical mass limit is reached.
The two scenarios for SN Ia formation differ in the nature
 of the companion.
This is a main sequence or red giant star in the
 so-called 'single degenerate'\ scenario, and is another
 white dwarf in the 'double degenerate'\ scenario.

The purpose of the 'Supernova Ia progenitor survey'\ (SPY) was to
 check the double degenerate scenario by observational means.
Thus, we observed more than 1000 WD over the course of four years at the ESO
 VLT equipped with UVES in order to check the objects for radial velocity
 (RV) variations (cf. Napiwotzki et al. 2003).
% which is an unambigous hint for binarity
Follow-up observations of promising objects were performed in order to derive
 system parameters like periods ($P$) and RV semi-amplitudes ($K$).
In combination with quantitative spectral analyses we computed
 the systems' total masses and merging times.
A very promising SN\,Ia precursor candidate was discovered 
 by Napiwotzki et al. (2005) 
 in the course of our project.

Due to mis-classification in the input catalogue the SPY sample also contains a
 number of subdwarf B stars (sdBs; Lisker et al. 2005).
Since these objects are immediate precursors of white
 dwarfs, they are also promising objects with respect to the search for SN Ia
 progenitors.
For example, Maxted et al. (2000) and Geier et al. (2006)
 found the sdB binary KPD\,1930+2752
 to be a SN Ia precursor candidate.
Thus promising RV variable sdB stars were included in our follow-up 
 observations as well.

In this paper we present the results of an analysis of six sdB
 binaries discovered by SPY.

\sectionb{\bf 2}{\bf OBSERVATIONS AND RADIAL VELOCITY CURVES}\\
All program stars were observed at least twice 
 in the course of the SPY project at the ESO\,VLT.
Additional observations were made during follow-up campaigns
 at the ESO\,NTT (equipped with EMMI),
 the ESO\,VLT (UVES), the Calar Alto Observatory 3.5m telescope
 (TWIN) and the 4m WHT (ISIS) at La Palma.
% Therefore, we got up to 50 observations for our program stars

Radial velocities of the individual observations are determined
 by calculating the
 shifts of the measured wavelengths relative to their laboratory values. 
Since the S/N ratio was of the order of 15 to 30,
 most of the narrow metal lines were
 hardly visible in our individual spectra.
We therefore focussed on
 all available He\,{\sc i} lines, and on
 the observed H${\alpha}$ line profile because of its
 sharp and well-defined non-LTE line core.
We performed a simultaneous fit of a set of mathematical functions to the
 observed line profiles using the ESO MIDAS package.
A linear function was used to reproduce the overall spectral trend,
 and a Gaussian for the innermost line core.
In order to fit the H${\alpha}$ profile we used an additional
  Lorentzian to model the broad line wings.
The central wavelength of the Lorentzian was fixed to that of the Gaussian
 for physical reasons.

% 
% The RV of each spectrum was measured by fitting
%  Gaussians to 
%  the observed H${\alpha}$ line profiles
%  and all available He\,{\sc i} lines.
% In addition, a linear function was used to reproduce
%  the continuum and a Lorentzian 
%  to model the broad H${\alpha}$ line wings.
% The RVs were determined 
%  with respect to the central wavelengths of the fitted 
%  Gaussians.

The period search was carried out by means of a periodogram
 based on the 'Singular Value Decomposition' \ method.
For a large range of periods the best fitting sine-shaped RV curve 
was computed (see Napiwotzki et
 al.~2001).
The difference between the observed radial velocities and the
 best fitting theoretical RV curve
 for each phase set was evaluated in terms of the logarithm of 
 the sum of the squared
 residuals ($\chi^2$) as a function of period.
This method finally results in the data-set's power spectrum which
 allows to determine
 the most probable period of variability 
 (see Lorenz et al.~1998).

From the best fit RV curve corresponding to the most probable period, the ephemeris, the
 system's velocity and the semi-amplitude were derived.
As an example, 
 Fig.~1 displays the resulting power spectrum and 
 best-fit sine curve for HE~0532$-$4503.
The ephemerides for all program stars, are given in 
 Tab.~1,
 as well as the semi-amplitudes ($K$) and the derived
 system velocities ($\gamma$).

% 
% Sinosoidal functions were fitted
%  for a range of periods,
%  yielding `power spectra'\ which 
%  indicate the quality of the sine\,-\,fits as
%  a function of period.
% Subsequently, we 
%  fitted the semi-amplitudes $K$,
%  the system velocities $\gamma$ and the ephemeriedes
%  $HJD{\rm (T_0)}$ using sine curves again,
%  having the periods fixed.
% As an example, 
%  Fig.~1 displays the resulting power spectrum and 
%  best-fit sine curve for HE~0532$-$4503.
% The ephemerides for all program stars, are given in 
%  Tab.~2,
%  as well as the semi-amplitudes $K$ and the derived
%  system velocities $\gamma$.
%  

% \begin{center}
% {\parbox{125mm}{
% {\bf Table 1.}{
%  Program stars: coordinates and magnitudes
% }}}\\
% \begin{tabular}{l|ccc}
% \tablerule
% System & RA (2000) & $\delta$ (2000) & V \\
%        & [hh:mm:ss.ss] & [dd:mm.ss.s] & [mag] \\
% \tablerule
% WD\,0048$-$202  & 00:51:03.97 & $-$20:00:00.3 & 14.85 \\
% HE\,0532$-$4503 & 05:33:40.51 & $-$45:01:35.3 & 16.02 \\
% HE\,0929$-$0424 & 09:32:02.15 & $-$04:37:37.8 & 15.33 \\
% HE\,1448$-$0510 & 14:51:13.13 & $-$05:23:16.9 & 14.42 \\
% HE\,2135$-$3749 & 21:38:44.18 & $-$37:36:15.1 & 13.77 \\
% HE\,2150$-$0238 & 21:52:35.81 & $-$02:24:31.6 & 15.91 \\
% \tablerule
% \end{tabular}
% \end{center}

\begin{center}
\vbox{
\centerline{\psfig{figure=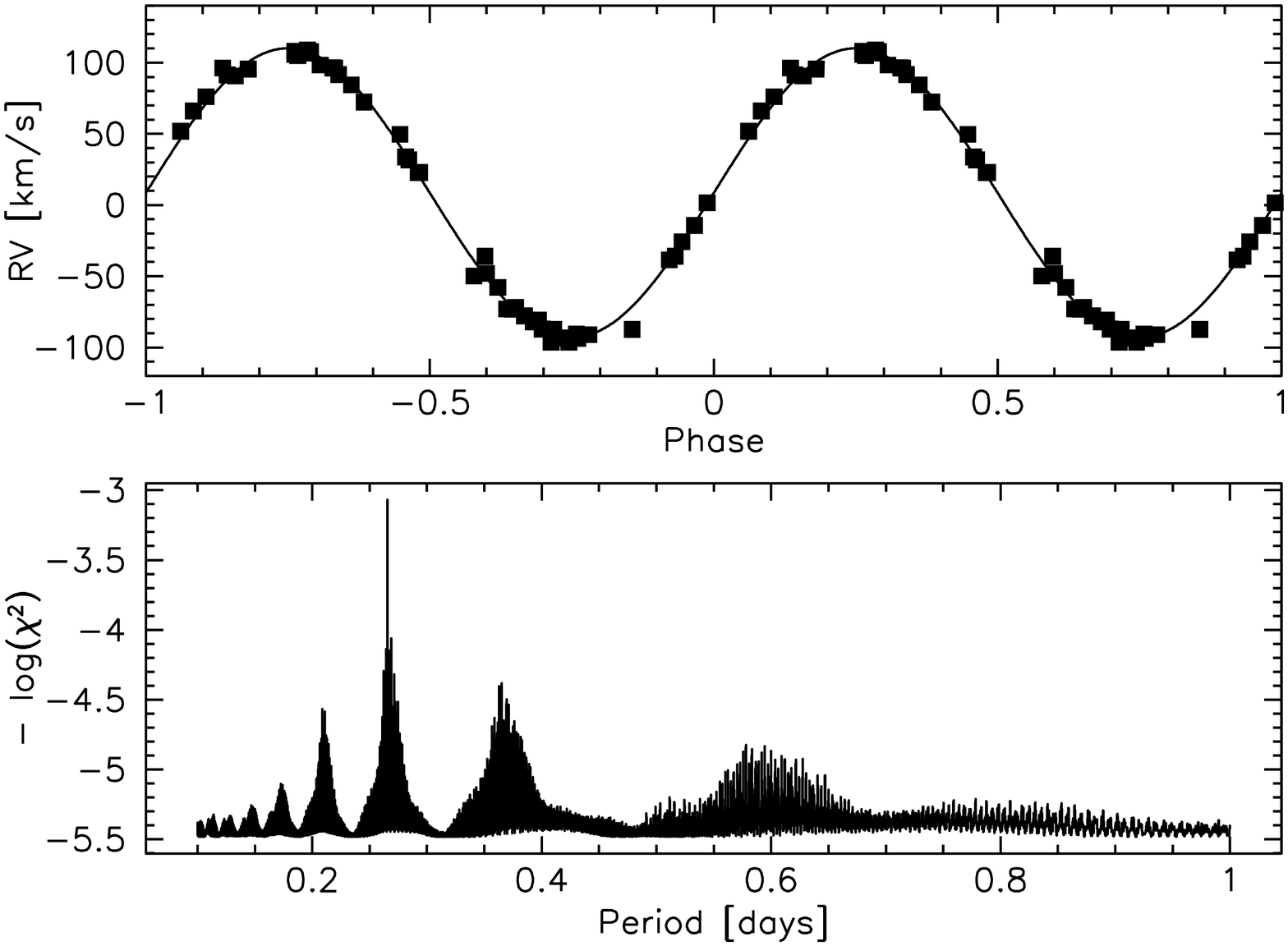,width=90truemm,angle=270}}
% \captionc{1}{
{\parbox{125mm}{
{\bf Figure~1.}{
 Sample best fit RV curve and power spectrum for the visible sdB star
 in the HE\,0532-4503 system.
Upper panel: Measured radial velocities as a
 function of orbital phase and fitted sine curve.
Lower panel: Power spectrum.
}}}
}
\end{center}

\begin{center}
\vbox{
{\parbox{125mm}{
{\bf Table 1.}{
 System parameters: ephemeris,
 RV semi-amplitudes $K$ and system velocities $\gamma$, for all the sdB stars discussed
}}}\\
\begin{tabular}{l|cr@{$\pm$}lr@{$\pm$}l}
\tablerule
System & Ephemeris & \multicolumn{2}{c}{$K$} & \multicolumn{2}{c}{$\gamma$} \\
       & [hel.JD$_{\rm 0}-$2\,450\,000] & \multicolumn{2}{c}{[km\,s$^{-1}$]} 
       & \multicolumn{2}{c}{[km\,s$^{-1}$]} \\
\tablerule
WD\,0048$-$202  & 3\,097.5596 $\pm$ 7.4436 $\times E$ &  47.9 & 0.4 & $-$26.5 & 0.4 \\
HE\,0532$-$4503 & 3\,099.9975 $\pm$ 0.2656 $\times E$ & 101.5 & 0.2 &     8.5 & 0.1 \\
HE\,0929$-$0424 & 3\,100.0585 $\pm$ 0.4400 $\times E$ & 114.3 & 1.4 &    41.4 & 1.0 \\
HE\,1448$-$0510 & 3\,097.0703 $\pm$ 7.1588 $\times E$ &  53.7 & 1.1 & $-$45.5 & 0.8 \\
HE\,2135$-$3749 & 3\,099.6520 $\pm$ 0.9240 $\times E$ &  90.5 & 0.6 &    45.0 & 0.5 \\
HE\,2150$-$0238 & 3\,100.6081 $\pm$ 1.3209 $\times E$ &  96.3 & 1.4 & $-$32.5 & 0.9 \\
\tablerule
\end{tabular}
}
\end{center}

\sectionb{\bf 3}{\bf QUANTITATIVE SPECTRAL ANALYSIS}\\
Prior to quantitative spectral analysis the spectra were corrected for the 
 measured RV and coadded in order to increase
 the S/N ratio.
Effective temperatures ($T_{\rm eff}$), surface gravities ($\log g$) and
 helium abundances ($\log\,[n_{\rm He}/n_{\rm H}]$) were
 determined by fitting simultaneously each observed hydrogen and helium line
 with a grid of metal-line blanketed LTE model spectra.
The procedure used is described in detail in Napiwotzki et al. (1999).
Because of its sensitivity to non-LTE effects, the H$\alpha$ line was
 excluded from this analysis.
Results are displayed in Tab.~2, and a sample
 fit is shown in Fig.~2.

\begin{figure}[t!]
\begin{center}
\vbox{
\centerline{\psfig{figure=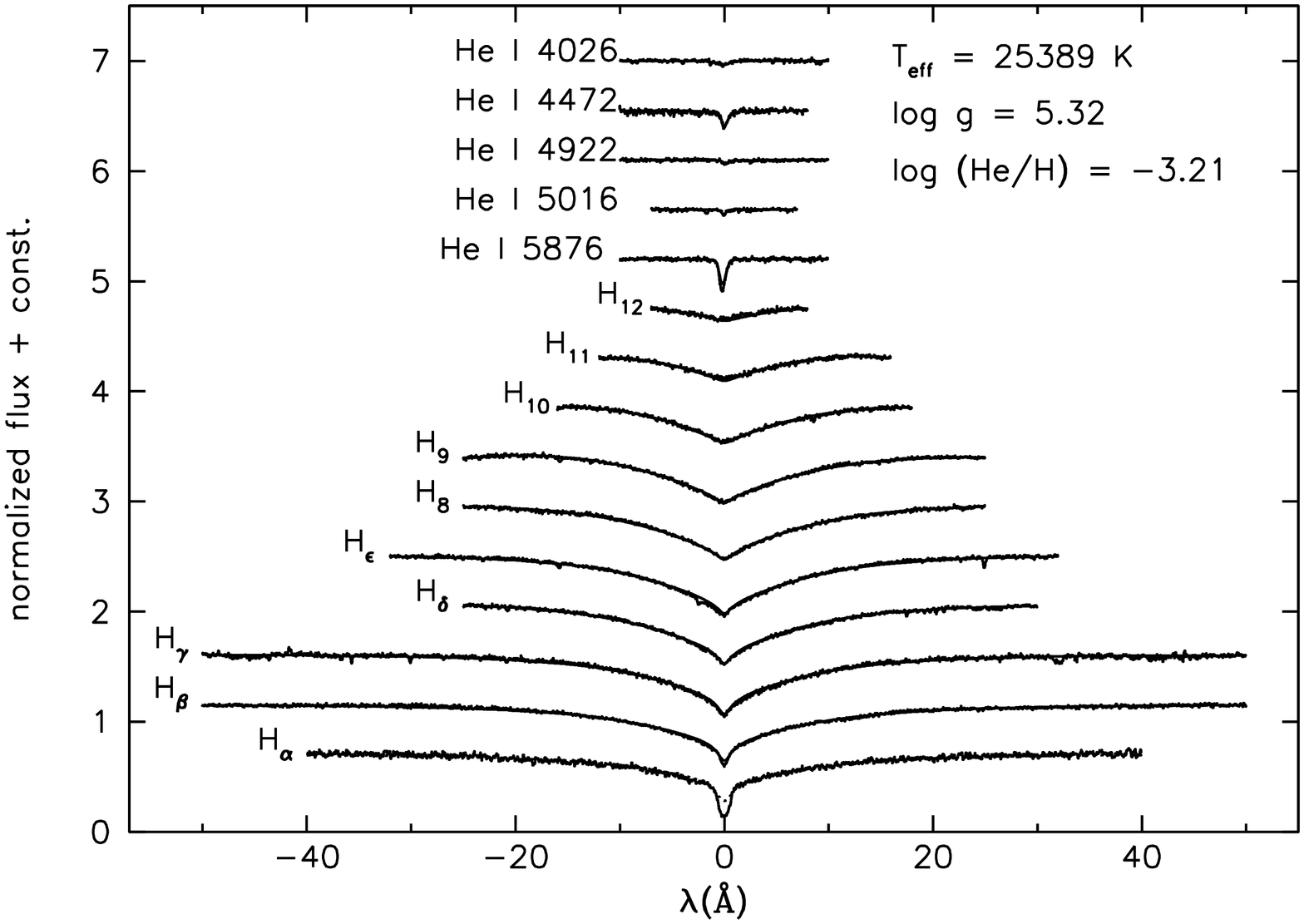,width=85truemm,angle=270}}
{\parbox{125mm}{
{\bf Figure~2.}{ 
 Sample model fit for a sdB star (HE\,0532-4503) based on
 47 UVES spectra.
}}}
}
\end{center}
\end{figure}

In addition, LTE metal abundances were derived for the
 program stars from measured equivalent widths using the classical
  curve-of-growth method.
Some stars show very rich abundance patterns with many species above the
detection threshold
 (e.g. HE\,2135$-$3749),
 while others like HE\,1448$-$0510 are extremely metal poor (see also 
Edelmann 2003).
In order to derive upper limits to elemental abundances in these cases,
 we assumed the detection limit
 for metal lines to be equal to the S/N level
 (in terms of equivalent width).
% Thus, we found an overabundance of Ti\{\sc iii} and Zn\,{\sc iii}
%  for the systems WD\,0048$-$202 and HE\,2135$-$3749.
Results of our metal abundance analysis are given in Tab.~3.

\begin{center}
\vbox{
{\parbox{125mm}{
{\bf Table 2.}{
 Stellar parameters:
 effective temperatures, surface gravities and helium abundances
 of the visible components.
Typical error margins for  $T_{\rm eff}$, $\log g$ and 
 $log\,[n_{\rm He}/n_{\rm H}]$ are
 235\,K, 0.03\,dex and 0.15\,dex, respectively.
}}}\\
\begin{tabular}{l|cccccc}
\tablerule
System & $T_{\rm eff}$ & & $\log g$ & \multicolumn{3}{c}{$\log [n_{\rm He}/n_{\rm H}]$} \\ % No of UVES spectra
       & [K]           & & [$\rm cms^{-2}$]    &                               \\
\tablerule
WD\,0048$-$202  & 29\,960 & & 5.50 & & $\le -$4.00 \\ % 9 
HE\,0532$-$4503 & 25\,390 & & 5.32 & & $-$3.21 \\ % 47 
HE\,0929$-$0424 & 29\,470 & & 5.71 & & $-$1.99 \\ %7
% HE\,1421$-$1206 & only 2x SPY -> Lisker et al.
HE\,1448$-$0510 & 34\,690 & & 5.59 & & $-$3.06 \\ % 3
HE\,2135$-$3749 & 30\,000 & & 5.84 & & $-$2.54 \\ % 14
HE\,2150$-$0238 & 30\,200 & & 5.83 & & $-$2.44 \\ % 2
\tablerule
\end{tabular}
}
\end{center}

\begin{center}
\vbox{
{\parbox{125mm}{
{\bf Table 3.}{ 
Metal abundance patterns of the program stars
 (relative to solar values)
}}}\\
\begin{tabular}{l|rrrrrr}
\tablerule
System & \multicolumn{4}{c}{Abundances $\epsilon$} \\
 & C\,{\sc ii}   & N\,{\sc ii}   & O\,{\sc ii}   & Mg\,{\sc ii}  & Al\,{\sc iii} & Si\,{\sc iii} \\
 & Si\,{\sc iv}  & S\,{\sc ii}   & S\,{\sc iii}  & Ar\,{\sc ii}  & Ar\,{\sc iii} & Ca\,{\sc iii} \\ 
 & Ti\,{\sc iii} & Fe\,{\sc iii} & Zn\,{\sc iii} \\ 
\tablerule
WD\,0048$-$202  
& \multicolumn{1}{c}{---} & $-$0.67 & $-$1.38 & $-$1.00 & $-$0.71 & $-$1.50 \\
& $-$1.98 & \multicolumn{1}{c}{---} & $-$1.41 & \multicolumn{1}{c}{---} & \multicolumn{1}{c}{---} & \multicolumn{1}{c}{---} \\
& +1.84 & $-$0.27 & \multicolumn{1}{c}{---} \\ \hline
HE\,0532$-$4503 
& $-$2.29 & $-$0.79 & $-$1.17 & $-$0.72 & \multicolumn{1}{c}{---} & $-$1.22 \\
& \multicolumn{1}{c}{---} & $-$0.28 & $-$0.82 & \multicolumn{1}{c}{---} & \multicolumn{1}{c}{---} & \multicolumn{1}{c}{---} \\
& \multicolumn{1}{c}{---} & 0.16 & \multicolumn{1}{c}{---} & \\ \hline
HE\,0929$-$0424 
& $-$1.81 & $-$0.14 & $-$1.07 & $-$0.80 & $-$0.53 & $-$0.93 \\
& \multicolumn{1}{c}{---} & \multicolumn{1}{c}{---} & $-$0.80 & +0.79 & \multicolumn{1}{c}{---} & \multicolumn{1}{c}{---} \\
& \multicolumn{1}{c}{---} & 0.00 & \multicolumn{1}{c}{---} & \\ \hline
HE\,1448$-$0510 
& $\le$$-$2.17 & $\le$$-$1.24 & $\le$$-$2.17 & $-$0.59 & $\le$$-$0.93 & \multicolumn{1}{c}{---} \\
& \multicolumn{1}{c}{---} & \multicolumn{1}{c}{---} & $\le$+0.80 & $\le$+0.67 & \multicolumn{1}{c}{---} & \multicolumn{1}{c}{---} \\
& \multicolumn{1}{c}{---} & $\le$$-$0.46 & \multicolumn{1}{c}{---} & \\ \hline
HE\,2135$-$3749 
& \multicolumn{1}{c}{---} & $-$0.36 & \multicolumn{1}{c}{---} & \multicolumn{1}{c}{---} & \multicolumn{1}{c}{---} & \multicolumn{1}{c}{---} \\
& \multicolumn{1}{c}{---} & $-$0.12 & $-$0.58 & +0.74 & +0.50 & +1.89 \\
& +1.61 & $-$0.87 & +1.94 & \\ \hline
HE\,2150$-$0238 
& \multicolumn{1}{c}{---} & $-$0.25 & $\le$$-$2.17 & $\le$$-$1.53 & $\le$$-$1.18 & $\le$$-$2.00 \\
& \multicolumn{1}{c}{---} & \multicolumn{1}{c}{---} & $-$0.26 & +0.92 & \multicolumn{1}{c}{---} & \multicolumn{1}{c}{---} \\
& \multicolumn{1}{c}{---} & $\le$$-$0.71 & \multicolumn{1}{c}{---} & \\
\hline
\end{tabular}
}
\end{center}

\sectionb{\bf 4}{\bf NATURE OF THE UNSEEN COMPANION}\\
Since the spectra of the program stars are single-lined,
 they reveal no information about the orbital motion of the
 sdBs' companions, and 
 thus we can only compute their mass functions
 \begin{equation}
 \label{equation-mass-function}
 f_{\rm m} = \frac{M_{\rm comp}^3 \sin^3i}{(M_{\rm comp} +
   M_{\rm sdB})^2} = \frac{P K^3}{2 \pi G} .
 \end{equation}
Although the RV semi-amplitude $K$ and the period $P$ are determined
 by the RV curve, $M_{\rm sdB}$, $M_{\rm comp}$ and $\sin^3i$ remain
 free parameters.
Binary population synthesis models (Han et al. 2003) indicate a most
 likely mass of $M_{\rm sdB}$\,=\,0.47\,M$_{\odot}$ for sdB stars,
 which we adopt for the following analysis.
Assuming $i$\,=\,90$^\circ$ we are able to compute the companions'
 minimum masses from Equation~\ref{equation-mass-function}.
The statistically most probable inclination angle
 is $i$\,=\,52$^\circ$
 which yields the most likely masses for the companions.
Our results are summarized in Tab.~4.

\begin{center}
\vbox{
{\parbox{125mm}{
{\bf Table 4.}{ 
 Mass functions, and masses
 (minimum mass $M_{\rm comp}^{\rm 90^{\circ}}$
 and most probable mass $M_{\rm comp}^{\rm 52^{\circ}}$)
 of the unseen companions.
}}}\\
\begin{tabular}{l|lcr}
\tablerule
System & $f_{\rm m}$ & $M_{\rm comp}^{\rm 90^{\circ}}$ &
 $M_{\rm comp}^{\rm 52^{\circ}}$ \\
 & [$M_{\odot}$] & [$M_{\odot}$] & [$M_{\odot}$] \\
\tablerule
WD\,0048$-$202  & 0.085 & 0.47 & 0.57 \\
HE\,0532$-$4503 & 0.029 & 0.25 & 0.37 \\
HE\,0929$-$0424 & 0.068 & 0.36 & 0.51 \\
HE\,1448$-$0510 & 0.115 & 0.56 & 0.68 \\
HE\,2135$-$3749 & 0.071 & 0.36 & 0.51 \\
HE\,2150$-$0238 & 0.122 & 0.48 & 0.70 \\
\hline
\end{tabular}
}
\end{center}

\sectionb{\bf 5}{\bf TIDALLY LOCKED ROTATION?}\\
For close binary systems, the components'
 stellar rotational velocities may be tidally locked to their
 orbital motions
 (see e.g.
 Napiwotzki et al. 2001), which means
\begin{equation}
v_{\rm rot} = \frac{2 \pi R_{\star}}{P} .
\end{equation}
Measurement of the projected rotational velocities 
 $v_{\rm rot}\,\sin\,i$ 
 would therefore allow to determine the 
 systems' inclination angles $i$.
In order to derive $v_{\rm rot}\,\sin\,i$,
 we compared the observed spectra 
 with rotationally broadened, synthetic line profiles.
The latter ones were computed for the stellar
 parameters given in Tab.~2 and
 Tab.~3.
Since sharp metal lines
 are much more sensitive to rotational broadening
 than Balmer or helium lines, we concentrated on
 strong N\,{\sc ii} lines between 5000 and 5008\,\AA.

\begin{figure}[t!]
\begin{center}
\vbox{
\centerline{\psfig{figure=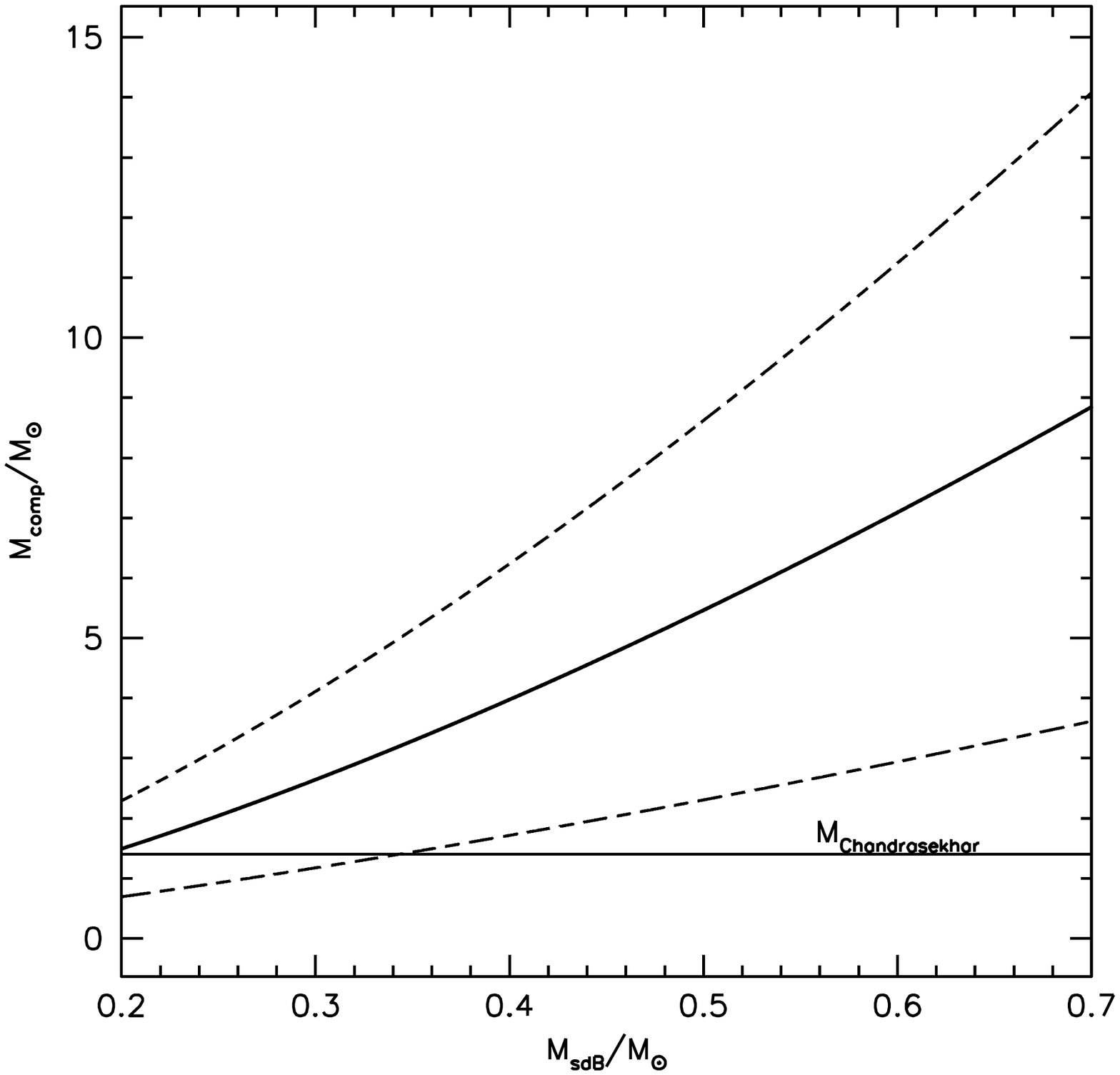,width=90truemm,angle=0}}
{\parbox{125mm}{
{\bf Figure~3.}{
 Determination of the mass of the unseen companion for the system
 HE\,0532-4503. 
 The solid line marks the mean mass for the unseen companion,
 while the dashed lines indicate the error margins.}
}}}
\end{center}
\end{figure}

Table~5 displays the resulting
 projected rotational velocities $v_{\rm rot}\,\sin\,i$, as well as 
 the deduced inclination angles $i$ and corresponding masses
 $M_{\rm comp}$.
No rotation velocity
 could be determined for HE\,1448$-$0510
 due to the lack of suitable metal lines,
 the remaining five stars are slow rotators with 
%  $v_{\rm rot}\,\sin\,i\,\le$\,10\,km\,s$^{-1}$.
 projected rotational velocities below 10\,km\,s$^{-1}$.
For the short period systems 
 HE\,0532$-$4503, HE\,0929$-$0424 and HE\,2135$-$3749
 we deduce inclination angles between 11$^\circ$ and
 32$^\circ$. 
The analyses of 
 WD\,0048$-$202 and HE\,2150$-$0238, however, yield 
 unreasonable values $\sin\,i\,>\,1$.
A possible explanation is, that the
 latter systems are NOT tidally locked, which is plausible
 because they have rather long periods.

\begin{center}
\vbox{
{\parbox{125mm}{
{\bf Table 5.}{
Periods and projected rotational velocities, as well as 
 inclination angles and masses of the unseen companion computed
 for tidally locked rotation.
}}}\\
\begin{tabular}{l|cccccc}
\tablerule
System & $P$ & $R_{\rm sdB}$ & $v_{\rm rot}$ & $v_{\rm rot}\,\sin\,i$ & $i$ & $M_{\rm comp}$  \\
       & [d] & [$R_{\odot}$] & [${\rm km\,s^{-1}}$] & [${\rm km\,s^{-1}}$] & [deg] & [$M_{\odot}$] \\ 
\tablerule
HE\,0532$-$4503 & 0.2656 $\pm$ 0.0001 & 0.25 & 47 & 9 $\pm$ 2 & 11 $\pm$ 3 & 5.0 $\pm$ 3.0 \\
HE\,0929$-$0424 & 0.4400 $\pm$ 0.0002 & 0.16 & 18 & 6 $\pm$ 2 & 19 $\pm$ 7 & 2.7 $\pm$ 2.0\\
HE\,2135$-$3749 & 0.9240 $\pm$ 0.0003 & 0.14 & 7  & 4 $\pm$ 2 & 32 $\pm$ 20 & 1.1 $\pm$ 0.7\\
HE\,2150$-$0238 & 1.3209 $\pm$ 0.0050 & 0.14 & --- & 8 $\pm$ 2 & $\sin i\,>$\,1 & ---\\
HE\,1448$-$0510 & 7.1588 $\pm$ 0.0130 & 0.18 & --- & --- & --- & ---\\
WD\,0048$-$202  & 7.4436 $\pm$ 0.0150 & 0.20 & --- & $\le$\,5 & $\sin i\,>$\,1 & --- \\
\hline
\end{tabular}
}
\end{center}

The masses deduced for the companions of the short-period systems
 are quite large, as can be seen from Tab.~5.
For the HE\,2135$-$3749 system,
 the sdB companion's mass is 1\,$M_{\odot}$ which indicates
 a massive white dwarf.
The companion of HE\,0929$-$0424
 is even more massive, but the large error in mass prevents us from drawing conclusions.
For HE\,0532$-$4503 the companion mass is larger than the Chandrasekhar mass, even is we allow for the large error and adopt the canonical mass for the sdB, indicating that it might possibly be a black hole. 
As can be seen from Fig.\,3, the companion mass could be sub-Chandrasekhar only if the sdB is of very low mass ($M_{\rm sdB} < 0.34M_{\rm \odot}$). As indicated by some tests it will be possible to reduce the uncertainty in $v_{\rm rot}\,\sin\,i$ if we include more lines. This will be done in the near future.

Black holes are rare objects and it is therefore very unlikely
 that our small sample of six RV variable sdBs contain two of them.
But stellar rotation and orbital motion
 may also be locked in a period-ratio
 that differs from unity (like Mercury, for which the ratio is 3/2).

Since the projected rotational velocities are small,
 the crucial parameter for the measurement
 of $v_{\rm rot}\,\sin\,i$ and subsequently
 for $M_{\rm comp}$ is
 the spectral resolution of the instrument.
Our spectra, however, have been measured through
 rather wide slits.
Their spectral resolution is therefore
seeing dependent. We used information from the seeing monitor to
determine the instrumental profile during these observations. When coadding 
the spectra we discarded those (few) taken in poor conditions and used only 
those taken under similar seeing conditions $\approx \, 1"$. This procedure may have 
led to an underestimation of the width of the instrumental profile which could 
have led to an overestimation of the system inclination. A lower inclination 
means even higher masses. Only if we should still have overestimated the 
width of the instrumental profile, the companion masses could be lower.  
To obtain more accurate values for $v_{\rm rot}\,\sin\,i$ we will
need additional observations. These will have to be taken using a
small slit so that the instrumental profile is well defined. For the time 
beeing the resulting high companion masses have to be taken with a pinch of 
salt.\\
\\

% 
% ACKNOWLEDGMENTS. Special thanks are to  John Brown and Fred White
% for their help.
% 

\goodbreak

\References
\refb Edelmann H., 2003, PhD thesis, University of Erlangen-Nuremberg
\refb Geier S., Heber U., Przybilla N., Kudritzki R. P., 2006, in Proc. of the Second Meeting on Hot Subdwarf  
 Stars, Baltic Astronomy
\refb Han Z., Podsiadlowski P., Maxted P.F.L., Marsh T.R., 2003, MNRAS
 341, 669
\refb Lisker T., Heber U., Napiwotzki R., et al., 2005, A\&A, 430, 223
\refb Lorenz L., Mayer P., Drechsel H., 1998, A\&A 332, 909
\refb Maxted P.F.L., Marsh T.R., North R.C., 2000, MNRAS 317, L41
\refb Napiwotzki R.,  Christlieb N., Drechsel H., et al., 2003,
 ESO Msngr 112, 25
\refb Napiwotzki R., Edelmann H., Heber U., et al., 2001,
 A\&A 378, L17 % HE\,1047$-$0436
\refb Napiwotzki R., Green P.J., Saffer R.A., 1999, ApJ 517, 399
\refb Napiwotzki R.,  Karl, C., Nelemans, G., et al., 2005, in Proc. of the
 9th European Workshop on White Dwarfs, ed. D. Koester and S. Moehler, 
 ASP conference series, Vol. 334

\vskip1mm

% \centerline{\fbox{\bf Citation of papers in workshop proceedings}}
% \refb
% Nitta~A., Winget~D.~E., Kanaan~A. et al. 1999, in
% {\it 11th European Workshop on White Dwarfs}, eds. J-E.~Solheim \& E.
% G.~Mei\v{s}tas, ASP Conf. Ser., 169, 144
% 
% 
% \centerline{\fbox{\bf Citation of papers in colloquium proceedings}}
% \refb
% Weiss~W.~W. 1986, in {\it Upper Main Sequence Stars with Anomalous
% Abundances}, IAU Colloq. 90, eds. C.~R.~Cowley, M.~M~Dworetsky \&
% C.~M\'{e}gassier, Reidel Publ. Company, Dordrecht, p. 219
%
%
% \centerline{\fbox{\bf Citation of papers in books}}
% \refb
% Searle~L., Sargent~W.~L.~W. 1968 in {\it The Magnetic and Related
% Stars}, ed. R. Cameron, Mono Book Corporation, Baltimore, p. 219

\end{document}